\documentclass[pdflatex,final,3p,times]{elsarticle}
\usepackage{axodraw}
\usepackage{color}
\bibstyle{elsarticle-num}

\def\SOFTSUSY{{\tt SOFTSUSY}}
\def\SUSPECT{{\tt SUSPECT}}
\def\SPHENO{{\tt SPHENO}}
\def\ISASUGRA{{\tt ISASUGRA}}

\newcommand{\newc}{\newcommand}
\newc{\half}{\frac{1}{2}}
\newc{\eps}{\epsilon}
\newc{\lam}{{\bf \lambda}}
\newc{\lamp}{{\bf \lambda}^{\prime}}
\newc{\lampp}{{\bf \lambda}^{\prime\prime}}
\newc{\nonr}{\nonumber}
\newc{\kap}{\kappa}
\newc{\lab}[1]{\label{eq:#1}}
\newc{\bino}{\widetilde{\cal B}}
\newc{\wino}{\widetilde{\cal W}}
\newc{\higgsino}{\tilde{h}}
\newc{\gluino}{\widetilde{\cal G}}
\newc{\code}[1]{{\tt #1}}
\newc{\ovl}{\overline}
\newc{\mlh}[1]{({m}_{\tilde{L}_{#1} H_1}^2)}
\newc{\mhl}[1]{({m}_{H_1 \tilde{L}_{#1}}^2)}
\newc{\ml}{{( m_{\tilde{L}}}^2)}
\newc{\rpc}{{\mbox{$R_p$}}}
\newc{\rpv}{{\mbox{${\not\!\!R_p}$}}}

\newc{\mtxt}[1]{{\color{blue}  #1}}
\newc{\stxt}[1]{{\color{red}  #1}}

\journal{Computer Physics Communications}

\begin{document}

\begin{frontmatter}

\title{Computation of Neutrino Masses in $R-$parity Violating Supersymmetry in \SOFTSUSY}

\author[damtp]{B.C.~Allanach}\author[damtp,cav]{C.H.~Kom}
\author[bonn]{M.~Hanussek}
\address[damtp]{DAMTP, CMS, University of Cambridge, Wilberforce road,
  Cambridge, CB3 0WA, United Kingdom}
\address[cav]{The Cavendish Laboratory, J J Thomson Ave, University of
  Cambridge, 
  Cambridge, CB3 0HE, United Kingdom}
\address[bonn]{Bethe Center of Theoretical Physics, University of Bonn, Nussallee 12, D-53115
  Bonn, Germany}
\begin{abstract}
  The program {\tt SOFTSUSY}~can calculate tree--level neutrino masses
  in the $R-$parity violating minimal supersymmetric standard model
  (MSSM) with real couplings. At tree--level, only one neutrino acquires a
  mass, in  
  contradiction with neutrino oscillation data.  Here, we describe an
  extension to the {\tt SOFTSUSY}~program which includes one--loop
  $R-$parity violating effects' contributions to neutrino masses and
  mixing. Including the one--loop effects refines the radiative
  electroweak symmetry breaking calculation, and may result in up to
  three massive, mixed neutrinos.  This paper serves as a manual to
  the neutrino mass prediction mode of the program, detailing the
  approximations and conventions used.
\end{abstract}

\end{frontmatter}

\section{Program Summary}
\noindent{\em Program title:} \SOFTSUSY{}\\
{\em Program obtainable
  from:} {\tt http://projects.hepforge.org/softsusy/}\\
{\em Distribution format:}\/ tar.gz\\
{\em Programming language:} {\tt C++}, {\tt fortran}\\
{\em Computer:}\/ Personal computer\\
{\em Operating system:}\/ Tested on Linux 4.x\\
{\em Word size:}\/ 32 bits\\
{\em External routines:}\/ None\\
{\em Typical running time:}\/ A second per parameter point.\\
{\em Nature of problem:}\/ Calculation of neutrino masses and the neutrino
mixing matrix at one--loop level
in the $R$--parity violating minimal supersymmetric standard 
model. The solution to the renormalisation group equations must be consistent
with a high or weak--scale boundary condition on supersymmetry breaking
parameters and $R$--parity violating parameters, as well as a weak--scale boundary condition
on gauge couplings, Yukawa couplings and the Higgs potential parameters.\\
{\em Solution method:}\/ Nested iterative algorithm. \\
{\em Restrictions:} {\SOFTSUSY} will provide a solution only in the
perturbative r\'{e}gime and it assumes that all couplings of the MSSM are real
(i.e.\ $CP-$conserving). 

\newpage

\section{Introduction}

Supersymmetric (SUSY) models provide an attractive weak--scale
extension to the Standard Model (SM).  The R--parity
conserving (\rpc) minimal supersymmetric extension of the Standard Model
(\rpc~MSSM) is often used as a reference model for phenomenological
studies.  There exist several publicly available spectrum generators
for the $\rpc$ MSSM: \ISASUGRA~\cite{Paige:2003mg},
\SOFTSUSY~\cite{Allanach:2001kg}, \SUSPECT~\cite{Djouadi:2002ze} and
\SPHENO\footnote{\SPHENO~includes a small subset of the $\rpv$ interactions.}~\cite{Porod:2003um}.  Spectrum information is typically
transferred to decay packages and event generators via a file in the
SUSY Les Houches Accord format~\cite{Skands:2003cj,Allanach:2008qq}.

The most general renormalisable superpotential of the MSSM contains
$R-$Parity violating ($\rpv$) couplings which violate baryon and lepton
number~\cite{Dreiner:1997uz}. A symmetry can be imposed upon the model
in order to maintain stability of the proton, for example baryon
triality~\cite{Ibanez:1991pr} or proton
hexality~\cite{Dreiner:2007vp}.  It has been shown that $\rpv$ models
may have interesting features, such as the generation of non--zero
neutrino masses and mixing without the addition of right--handed neutrino
fields 
\cite{Hall:1983id}.  In fact, neutrino oscillation data indicates that
at least two neutrinos must be massive, so a realistic extension to
the SM should include mechanisms to generate these masses.  Here, we
describe an extension to \SOFTSUSY\@ which calculates neutrino masses
and mixing to one--loop order in the presence of $\rpv$ couplings.
The latest version of \SOFTSUSY~including $\rpv$ effects can be
downloaded from the address
\begin{quote}
{\tt http://projects.hepforge.org/softsusy/}
\end{quote}
Installation instructions and more detailed technical documentation of the
code may also be found there.

The $\rpc$ and $\rpv$ aspects of the \SOFTSUSY~calculation leading to
self--consistent spectra are already explained in detail in
Refs.~\cite{Allanach:2001kg} and~\cite{Allanach:2009bv} respectively,
with the technical differences between these two calculations detailed
in the latter reference.  They shall not be repeated here.  Instead we shall
concentrate on the 
calculation of neutrino masses and mixing, and differences in the
$\rpv$ calculations between the new and the previous release, which
are called the neutrino mode and the $\rpv$ mode respectively.  
Two new features of {\tt SOFTSUSY3.2}~are the $\rpv$ one--loop tadpole
corrections to the 
two Higgs vacuum expectation values (VEVs), and the complete one--loop
tadpole corrections to the sneutrino VEVs.  
These improvements allow the computation of the neutrino spectrum and neutrino
mixings, with minimal additional computational cost. They also slightly change
some predictions in the non-neutrino sector. 

We proceed with a definition of the \SOFTSUSY~convention for the
$\rpv$ parameters and mixings relevant for the neutrino mass
calculation in section~\ref{sec:notation}. Next, in
section~\ref{sec:calculation}, we discuss the calculation of the
neutrino masses, detailing the approximations made.  Installation
instructions can be found on the \SOFTSUSY~web--site, but instructions
to run the program can be found in~\ref{sec:run}.  The output
from a \SOFTSUSY~sample run in the neutrino mode is displayed and
discussed in~\ref{sec:output}, whereas a sample main program
is shown and explained in~\ref{sec:prog}. Some more technical
information on the structure of the program can be found in~\ref{sec:objects}.
It is expected that the information in~\ref{sec:objects} will only be of use
to users who wish to 
`hack' \SOFTSUSY~in some fashion.

\section{MSSM $\rpv$ Parameters \label{sec:notation}}

In this section, we introduce the $\rpv$ MSSM parameters in the
\SOFTSUSY~conventions relevant for the neutrino mass and 1--loop
tadpole calculations.  A detailed description of the complete set of
$\rpv$ MSSM parameters is presented in Ref.~\cite{Allanach:2009bv}.
The latter follows Ref.~\cite{Allanach:2003eb} and so the notation and
conventions employed are similar. 

\subsection{$\rpv$ supersymmetric and SUSY breaking parameters \label{susypars}}

The chiral superfield particle content of the MSSM has the following
$SU(3)_c\times SU(2)_L\times U(1)_Y$ quantum numbers:
\begin{eqnarray}
L:&(1,2,-\half),\quad {\bar E}:&(1,1,1),\qquad\, Q:\,(3,2,\frac{1}{6}),\quad
{\bar U}:\,(\bar 3,1,-\frac{2}{3}),\nonr\\ {\bar D}:&(\bar 3,1,\frac{1}{3}),\quad
H_1:&(1,2,-\half),\quad  H_2:\,(1,2,\half).
\label{fields}
\end{eqnarray}
$L$, $Q$, $H_1$, and $H_2$ are the left handed doublet lepton and
quark superfields and the two Higgs doublets. $\bar E$, $\bar U$, and
$\bar D$ are the lepton, up--type quark and down--type quark
right--handed superfield singlets, respectively.  The $\rpv$ part of
the renormalisable MSSM superpotential that violates lepton number,
written in the interaction eigenbasis, is
\begin{equation} 
W_{\rpv}\subset\eps_{ab}\left[ \frac{1}{2} \lam_{ijk} L_i^a L_j^b{\bar E}_k
  + \lamp_{ijk} L_i^a Q_j^{xb} {\bar D}_{kx} - \kap_i L_i^a H_2^b
  \right]\,.
 \label{superpot1} 
\end{equation} 
Here, $\kap_i$ [$\lam_{ijk}$ and $\lamp_{ijk}$] are the bi--linear
[trilinear] couplings.  We denote an $SU(3)$ colour index of the
fundamental representation by $\{x,y,z\} \in \{1,2,3 \}$. The
$SU(2)_L$ fundamental representation indices are denoted by $\{a,b,c\}
\in \{1,2\}$ and the generation indices by $\{i,j,k\} \in \{1,2,3\}$.
$\epsilon_{ab}=\epsilon^{ab}$ is a totally antisymmetric tensor, with
$\epsilon_{12}=1$.  Currently, only real couplings in the
superpotential and Lagrangian are included.

The corresponding soft $\rpv$ breaking interaction potential, together
with the bi--linear mixing term between the scalar component of the
lepton doublets and the $H_1$ superfields, is involved in tadpole and
1--loop corrections to the neutrino--neutralino masses.  It is given
by
\begin{equation}
  V_{\rpv} \subset \eps_{ab} \left[\frac{1}{2} h_{ij k}\tilde{L}_i^a
    \tilde{L}_j^b \tilde{{e}}_k +h^\prime_{i jk} \tilde{L}_i^a
    \tilde{Q}_j^{bx} \tilde{d}_{kx} -D_i \tilde{L}_i^a
    H_2^b~+~{\rm H.c.}\right] + \left[m^2_{\tilde L_i
      H_1}\tilde{L}^\dagger_{ia}H_1^a +~{\rm H.c.}\right]\,,
  \label{soft}
\end{equation}
where fields with a tilde are the scalar components of the superfield
with the same capital letter.  The scalar components
of the superfields $H_1$ and $H_2$ and the superfields themselves have
the same notation.  The electric charges of $\tilde d$ and $\tilde e$
are $\frac{1}{3}$, and $1$, respectively.  $h_{ijk}$ and
$h^\prime_{ijk}$ are trilinear soft SUSY breaking parameters that
correspond to $\lam_{ijk}$ and $\lamp_{ijk}$, and $D_i$ and
$m^2_{\tilde{L}_iH_1}$ are bi--linear SUSY breaking parameters.  ``H.c.''
denotes the Hermitian conjugate of the preceding terms.

\subsection{Neutral and charged fermion masses \label{sec:tree}}

In the presence of $\rpv$ interactions that violate lepton number, the
neutrinos and neutralinos mix, and the charginos and leptons mix
with each other.  Sneutrino--anti-sneutrino mixing is also present in
principle; in practice this has been shown to have negligible
phenomenological consequences once experimental bounds have been
applied~\cite{Dedes:2007ef}, and is neglected in our calculation.

The  (7$\times$7) neutrino--neutralino Lagrangian mass term, containing
three families of neutrinos is given in \cite{Allanach:2003eb} and reads
\begin{eqnarray}
{\cal L}= -\frac{1}{2} \; (\nu_i, -i \bino, -i \wino^{(3)}, \higgsino_1^0, \higgsino_2^0
)
\; {\cal M}_{\rm N} \; \left ( \begin{array}{r@{}l}
&\nu_j \\ -i&\bino \\  -i&\wino^{(3)} \\ &\higgsino_1^0 \\
&\higgsino_2^0 \end{array} \right ), \label{basisvector}
\end{eqnarray}
where at tree level,
\begin{eqnarray}
{\cal M}_{\rm N} = \left ( \begin{array}{ccccc}
 0_{ij}& -\frac{g'}{2} v_i   &  \frac{g_2}{2} v_i   &  0   & - \kappa_i \\[4mm]
 -\frac{g'}{2} v_j    &   M_1   &   0  &  -\frac{g'}{2} \langle H_1^0 \rangle   &  \frac{g'}{2} \langle H_2^0 \rangle \\[4mm]
  \frac{g_2}{2} v_j   &   0  &   M_2   &  \frac{g_2}{2} \langle H_1^0 \rangle   &  -\frac{g_2}{2} \langle H_2^0 \rangle\\[4mm]
 0    &   -\frac{g'}{2} \langle H_1^0 \rangle  &  \frac{g_2}{2} \langle H_1^0 \rangle    &   0  & -\mu \\[4mm]
 - \kappa_i    &  \frac{g'}{2} \langle H_2^0 \rangle   &  -\frac{g_2}{2} \langle H_2^0 \rangle   & -\mu    & 0 \\[4mm]
\end{array} \right ).\label{neutralino}
\end{eqnarray}
In Eq.~(\ref{basisvector}), the flavour basis neutral fermions are the
3 neutrinos ($\nu_i$), the gauginos ($\bino$,$\wino$) and the
higgsinos ($\higgsino_1^0$,$\higgsino_2^0$).  In the mass matrix in
Eq.~(\ref{neutralino}), $\kappa_i$ [$\mu$] are the supersymmetric
$\rpv$ [$\rpc$] bi--linear mixing parameters, $v_i$ [$\langle H_2^0 \rangle$ and $\langle H_1^0 \rangle$]
are the sneutrino [$H^0_2$ and $H^0_1$] vacuum expectation values
(VEVs), $M_1$, $M_2$ are the gaugino masses of hypercharge and weak
isospin respectively, and $g'$, $g_2$ are the corresponding gauge
couplings.  The mass eigenstates are obtained upon diagonalisation of
${\mathcal M}$: $\nu_{i=1,2,3},\tilde{\chi}^0_{1,2,3,4}$ via a 7 by 7
orthogonal matrix $O$:
\begin{equation}
{\mathcal M}_{\rm N}^{diag} = O^T {\mathcal M}_{\rm N} \; O. \label{Nmix}
\end{equation}
A simple multiplication of rows of $O$ by factors of $i$ can absorb
any minus signs in ${\mathcal M}_{\rm N}^{diag}$.  
Mass eigenstates are ordered in increasing (absolute) mass
eigenvalues.

To facilitate comparisons with neutrino oscillation data, it is useful
to express ${\cal M}_{\rm N}$ as
\begin{eqnarray}\label{neutralino34}
  {\cal M}_{\rm N} &=& \left( \begin{array}{cc} m_{\nu} & m \\ m^T &
    \mathcal{M}_{\chi^0}
  \end{array} \right),
\end{eqnarray}
where $m_{\nu}$ is a $3\times 3$ mass matrix, $\mathcal{M}_
{\tilde{\chi}^0}$ is a $4\times 4$ mass matrix, and $m$ is a $3\times
4$ matrix that mixes the neutrinos and the neutralinos. We define an
effective $3\times3$ neutrino mass matrix ${\cal M}_{\nu}$ via the
see--saw relation
\begin{eqnarray}\label{nueffective} 
  {\cal M}_{\nu} &\equiv& m_{\nu} - m \;
  \mathcal{M}_{\tilde{\chi}^0}^{-1} \; m^T.
\end{eqnarray}
It can be diagonalised by a $O(3)$ matrix
$O_{\nu}$,
\begin{equation}
{\mathcal M}_{\nu}^{diag} = O_{\nu}^T {\mathcal M}_{\nu}\, O_{\nu}. \label{Numix}
\end{equation}
For realistic mass spectra, $O_{\nu}$ is practically the same as
the $3\times3$ neutrino part of $O$ (we do not assume this, however). 

At tree level, the effective neutrino mass matrix is~\cite{Allanach:2003eb}
\begin{eqnarray}
{\cal M}_{\nu}^{\rm tree} &=& \frac{\mu (M_1g^2_2 + M_2g'^2)}{2 \langle H_1^0
  \rangle \langle H_1^0 \rangle (M_1g^2_2 + M_2g'^2)- 2 \mu M_1M_2}
\left( \begin{array}{ccc}
\Delta_1\Delta_1 & \Delta_1\Delta_2 &\Delta_1\Delta_3\\
\Delta_2\Delta_1 & \Delta_2\Delta_2 &\Delta_2\Delta_3\\
\Delta_3\Delta_1 & \Delta_3\Delta_2 &\Delta_3\Delta_3 \end{array} \right)\, ,
\label{mnutree_matrix}
\end{eqnarray}
where
\begin{eqnarray}
\Delta_i &\equiv&  v_i - \langle H_1^0 \rangle \frac{\kappa_i}{\mu}, \qquad i=1,2,3 \,.
\label{Lambda}
\end{eqnarray}
${\cal M}_{\nu}^{\rm tree}$ is rank 1 and so contains two zero eigenvalues.
Because of the presence of two massless neutrinos in the
spectrum, the tree level values are not realistic.  The numerical
values of $O_{\nu}$ are also not meaningful in the tree-level approximation.
The 1--loop corrections 
to ${\cal M}_{\nu}$ can lead to two or three non--zero neutrino
masses, if more than one lepton flavour is violated.  

By default, the
neutrino masses are normal--ordered,
i.e. $|m_{\nu_1}|<|m_{\nu_2}|<|m_{\nu_3}|$.  A spectrum with inverted
ordering can be obtained by
\begin{eqnarray}
\left ( \begin{array}{c}
\nu_1 \\ \nu_2 \\ \nu_3 \end{array} \right ) &\to&
\left ( \begin{array}{c}
\nu'_1 \\ \nu'_2 \\ \nu'_3 \end{array} \right ) = 
\left ( \begin{array}{c}
\nu_2 \\ \nu_3 \\ \nu_1 \end{array} \right ),
\end{eqnarray}
together with the corresponding swaps (of the column vectors) in the
mixing matrix $O_{\nu}$.  The mass ordering then becomes
$|m_{\nu'_3}|<|m_{\nu'_1}|<|m_{\nu'_2}|$.

The presence of $\rpv$ interactions also mix charged--leptons with the
charginos. The Lagrangian contains the ($5 \times 5$) chargino--lepton
mass matrix
\begin{eqnarray}
{\cal L}= - ( -i \wino^-, \higgsino_{1}^-, e_{L_j}^-
)
\; {\cal M}_{\rm C}\; \left ( \begin{array}{r@{}l}
 -i & \wino^+\\ &\higgsino_2^+ \\ &e_{R_k}^+ \\
 \end{array} \right ) + {\rm H.c.}\,.
\end{eqnarray}
The mass eigenstates $\ell=(e,\mu,\tau), \tilde{\chi}^\pm_{1,2}$ are
given upon the diagonalisation of the matrix ${\mathcal M}_{\rm C}$, where at
tree level,
\begin{eqnarray}
{\cal M}_{\rm C} = \left ( \begin{array}{ccc}
M_2  & \frac{g_2}{\sqrt{2}}\langle H_2^0 \rangle & 0_j  \\
 \frac{g_2}{\sqrt{2}}\langle H_1^0 \rangle  & \mu & -\frac{1}{\sqrt{2}}(Y_E)_{ij}v_i \\
\frac{g_2}{\sqrt{2}}v_i  & \kappa_i & \frac{1}{\sqrt{2}}\left((Y_E)_{ij} \langle H_1^0 \rangle+\lambda_{kij}v_k\right)\\
 \end{array}
\right ) , \label{chargino}
\end{eqnarray}
Here, $Y_E$ is the lepton Yukawa matrix from the $\rpc$
superpotential in Ref.~\cite{Allanach:2001kg}. We define the diagonalised mass
matrix 
\begin{eqnarray}
{\cal M}^{diag}_{\rm C}= U {\cal M}_{\rm C} V^T, \label{diagUV}
\end{eqnarray}
$U$ and $V$ being orthogonal 5 $\times$ 5 matrices.  
The one--loop $\rpv$ corrections to ${\cal M}_{\rm C}$ are related to neutrino
masses. For $\rpv$ effects giving small enough neutrino masses to pass
empirical bounds, the one--loop $\rpv$ corrections to ${\cal M}_{\rm C}$ are
expected to be negligible, unless one assumes that the $\rpv$ effects are
large, but have a high degree of cancellation in the neutrino masses. 
At present, the one--loop
$\rpv$ corrections to ${\cal M}_{\rm C}$ are not implemented.  The
  absolute mass eigenvalues are in increasing order along the diagonal
  of ${\cal M}^{diag}_{\rm C}$.  An effective charged lepton mixing
matrix, $U_l$, can be obtained from the 3 $\times$ 3, top right
part of $U$.  The PMNS matrix $U_{\rm PMNS}$ is then defined as
\begin{equation}\label{eq:upmns}
U_{\rm PMNS} = U_{l}^*O_{\nu},
\end{equation}
with $O_{\nu}$ including the (column) swapping when an inverted mass
ordering is desired.  

\subsection{Neutral slepton masses}
Lepton number violating interactions mix the sleptons and higgs'. In the limit
of CP conservation, as assumed here, one obtains two mass squared matrices:
one for the neutral CP even real scalars ${\mathcal M}^2_{\varphi^0_+}$, and
one for neutral CP 
odd real scalars ${\mathcal M}^2_{\varphi^0_-}$.  
The Lagrangian containing these terms looks like
\begin{eqnarray}\label{eq:mcp}
{\mathcal L}= -\frac{1}{2}(h^0_{2\pm} , h^0_{1\pm}, \tilde{\nu}_{j\pm})
\; {\mathcal M}^2_{\varphi^0_\pm}\; \left ( \begin{array}{c}
 h^0_{2\pm} \\ h^0_{1\pm} \\ \tilde{\nu}_{k\pm} \\
 \end{array}\right)\,,
\end{eqnarray}
where the fields with subscript $+(-)$ are the real (imaginary) part
of the complex fields in obvious notation.  At tree level,
\begin{eqnarray}\label{eq:mcptree}
  {\mathcal M}^2_{\varphi^0_\pm} &=&
  \frac{(g'^2+g_2^2)}{4}\left\{\left( \begin{array}{ccc} \langle H_2^0
      \rangle^2 & -\langle H_2^0 \rangle \langle H_1^0 \rangle &
      -\langle H_2^0 \rangle v_k \\ -\langle H_2^0 \rangle \langle
      H_1^0 \rangle & \langle H_1^0 \rangle^2 & \langle H_2^0 \rangle
      v_k \\ -\langle H_2^0 \rangle v_j & \langle H_1^0 \rangle v_j &
      v_j v_k \end{array} \right)-\frac{\left(\langle H_2^0 \rangle^2
      - \langle H_1^0 \rangle^2 - v_i^2 \right)}{2}
    \left( \begin{array}{ccc} 0 & 0 & 0 \\ 0 & 1 & 0 \\ 0 & 0 &
      \delta_{jk}\end{array} \right)\right\} \nonumber\\ && +
  \left( \begin{array}{ccc} \frac{m_3^2\langle H_1^0 \rangle + D_i
      v_i}{\langle H_2^0 \rangle} & \mp m_3^2 & \mp D_k \\ \mp m_3^2 &
    \mu^2+m^2_{H_1} & \mu \kappa_k + m^2_{\tilde{L}_kH_1}\\ \mp D_j &
    \mu \kappa_j + m^2_{\tilde{L}_jH_1} & \kappa_j\kappa_k +
    (m^2_{\tilde{L}})_{jk}\end{array} \right)\,,
\end{eqnarray}
where $m_3^2$ is the soft breaking term corresponding to the bilinear Higgs-mixing parameter $\mu$.


Currently, ${\mathcal M}^2_{\varphi^0_\pm}$ is calculated at tree-level and is
only used to calculate the one-loop corrections to neutrino masses and
mixings. 
The mass eigenstates are ordered in mass $(m_{\varphi^0_\pm})_{i=1,\ldots,5}$
\begin{equation}
  \label{eq:scalMix}
  \mbox{diag} (m_{\varphi^0_\pm})_i^2 = O_\pm^T {\mathcal M}^2_{\varphi^0_\pm}
  O_\pm 
\end{equation}
with mixing matrix $O_\pm$. 

\section{Calculation Algorithm \label{sec:calculation}}

\begin{figure}
\begin{center}
\begin{picture}(323,245)
\put(10,0){\makebox(280,10)[c]{\fbox{7.\ Calculate Higgs,
      sparticle \textbf{and neutrino} pole masses. Run to $M_Z$.}}}
\put(10,40){\makebox(280,10)[c]{\fbox{6.\ Run to $M_Z$.}}}
\put(150,76.5){\vector(0,-1){23}}
\put(10,80){\makebox(280,10)[c]{\fbox{5.\ Run to $M_X$. Apply soft breaking
and $\rpv$ SUSY boundary conditions.}}}
\put(150,116.5){\vector(0,-1){23}}
\put(10,130){\makebox(280,10)[c]{\fbox{\shortstack{4.\ EWSB, iterative solution of $\mu$ 
and sneutrino VEVs \\ \textbf{including full $\rpc$ and $\rpv$ 1 loop contributions to the
 tadpoles  $\frac{\partial {\bf \Delta V}}{\partial v_{u,d,i}}$ }.}}}}
\put(150,176){\vector(0,-1){23}}
\put(10,180){\makebox(280,10)[c]{\fbox{3.\ Run to $M_S$.}}}
\put(20,190){convergence}
\DashLine(115,185)(-45,185){5}
\DashLine(-45,185)(-45,5){5}
\DashLine(-45,5)(-10,5){5}
\put(-10,5){\vector(1,0){2}}
\put(150,217){\vector(0,-1){24}}
\put(150,259){\vector(0,-1){26}}
\put(60,265){\fbox{1.\ SUSY radiative corrections to
$g_i(M_Z)$}}
\put(10,220){\makebox(280,10)[c]{\fbox{2.\ $Y_E$, $Y_D$ include $v_k$
      contributions. Iterative solution of $Y_E$.}}} 
\put(182,45){\line(1,0){161}}
\put(343,45){\line(0,1){225}}
\put(343,270){\vector(-1,0){115}}
\end{picture}
\end{center}
\caption{Iterative algorithm used to calculate the $\rpv$ MSSM spectrum 
in the neutrino mode. Differences to the original $\rpv$ mode are highlighted in bold print.
The initial step is the
uppermost one. $M_S$ is the scale at which the EWSB
conditions 
are imposed, as discussed in the text. $M_X$ is the scale at which the high
energy SUSY breaking boundary conditions are imposed.
\label{fig:algorithm}}
\end{figure}
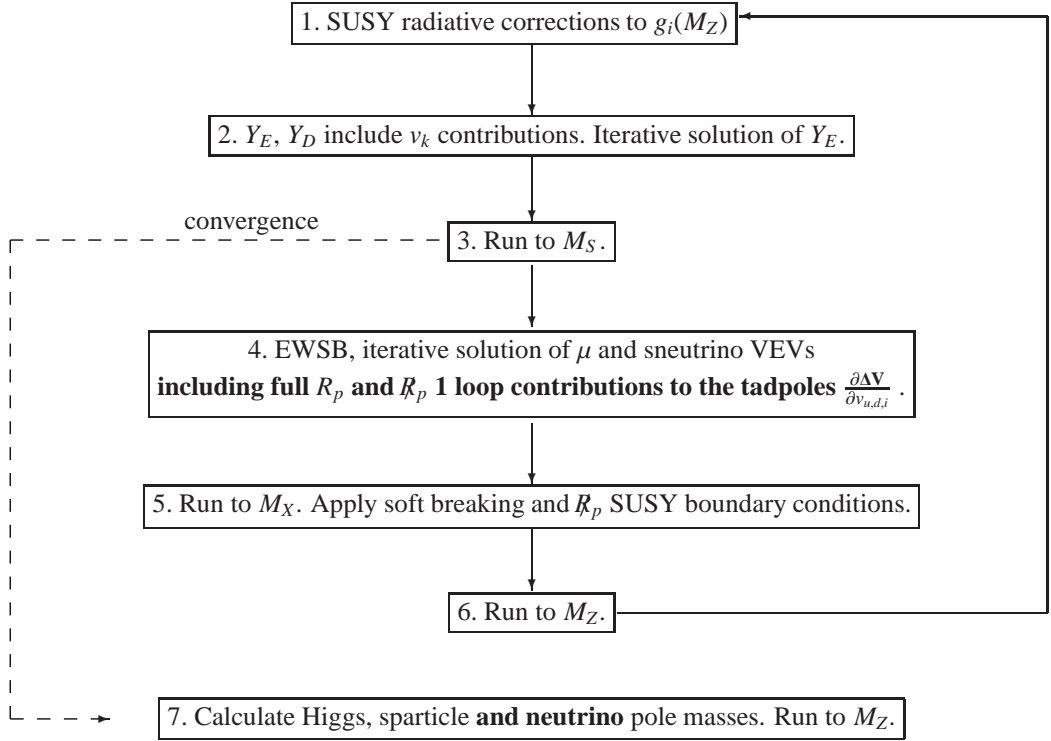

The calculation to calculate the low energy 
$\rpv$ mass spectrum proceeds mostly in the same way as the original $\rpv$
mode, which is detailed in Ref.~\cite{Allanach:2009bv}, but now also includes the
 full 1--loop $\rpv$ tadpoles and additional
methods to calculate the 1--loop 7x7 neutralino--neutrino mass matrix 
and to obtain the neutrino pole masses and the
PMNS matrix via the seesaw--mechanism. The calculation algorithm is displayed
in Fig.~\ref{fig:algorithm}, where the changes with respect to the $\rpv$ mode are displayed 
in bold print.
For a given a set of inputs for $\rpv$ couplings, assumptions about the
measurable quark mixing make a physical difference to the results. It is well
known that in the $R_p$ MSSM, a $U(3)^5$ global family symmetry,
where Yukawa couplings also transform as spurions,
renders
physical results invariant to where the CKM quark mixing lies. 
For example, the physics is identical for the cases where CKM mixing lies
solely in left handed up quarks and for the case where it lies solely in
left handed down quarks, since
the family symmetry relates the two cases~\cite{aaa}.  
However, in the $\rpv$ MSSM, one would {\em also}\/ have to transform 
the $\rpv$ Yukawa couplings $\lam'_{ijk}$ and $\lam''_{ijk}$ to keep the
theory invariant under the transformation. Since this is inconvenient, we wish
to leave the $\rpv$ couplings untransformed, and instead specify (in the basis
in which the $\rpv$ couplings are given), where the quark mixing lies. 
There is thus a physical difference depending on whether one assumes that CKM
mixing 
lies e.g.\ solely in the down--type quarks or solely in the up--type quarks. We
describe  the parameter in the program 
which specifies this mixing in~\ref{sec:objects}. It can have a significant
effect on the neutrino mass 
predictions. 

\subsection{$\rpv$ tadpole contributions \label{tadpole}}

The 1--loop $\rpv$ contributions to the tadpoles $\frac{\partial {\bf
    \Delta V}}{\partial v_{u,d}}$, as well as $\rpc$ and $\rpv$
contributions to $\frac{\partial {\bf \Delta V}}{\partial v_{i}}$, are
implemented in the neutrino mode of \SOFTSUSY~. These calculations include 
all three SM fermion families in the loop.  In the presence of $\rpv$
interactions, the charged 
sleptons and down--type Higgs mix, and the neutral scalars and Higgs mix with
each other, so out computation treats these contributions on an equal
footing.   
The sneutrino VEVs are now also 
calculated at 1--loop level at the renormalisation scale $M_S \equiv
\sqrt{m_{\tilde t_1}(M_S) \,m_{\tilde t_2}(M_S)}$.  
However,there are no renormalisation group equations (RGEs) implemented for
the sneutrino VEVs, so 
they remain constant at any renormalisation scale. 

\subsection{Neutrino--neutralino mass corrections}

The physical neutrino--neutralino masses are calculated to full
one--loop order, including both $\rpc$ and $\rpv$ contributions,
following Refs.~\cite{Dedes:2006ni,Allanach:2007qc} closely.  The
running parameters are evaluated at the renormalisation scale
$\mu=M_{SUSY}$. However, note that the physical neutralino masses
are displayed in the $\rpc$ limit in the \SOFTSUSY~output
without including the $\rpv$ contributions.  
However, in the absence of large cancellations, $\rpv$ corrections to
neutralino masses will be of order
the neutrino mass, i.e.\ negligible. 
To calculate the physical neutrino masses,
the external momentum for the physical neutrino--neutralino mass
matrix, ${\cal M}_{\rm N}^{phys}$, is set at $p=0$.  The matrix ${\cal
  M}_{\rm N}^{phys}$ is given by
\begin{equation}
  {\cal M}_{\rm N}^{phys} = {\cal M}_{\rm N} +
  \frac{1}{2}\left(\Delta_{\rm N} + \Delta^T_{\rm N}\right),
\end{equation}
where
\begin{equation}
  \Delta_{\rm N} = \Sigma_D - {\cal M}_{\rm N}\Sigma_L.
\end{equation}
In the above expression, $\Sigma_D$ and $\Sigma_L$ are mass
corrections and wavefunction renormalisation respectively.  The
effective physical neutrino mass matrix ${\cal M}_{\nu}$ is obtained
as defined in Eq.~\ref{nueffective}, from which the physical neutrino
masses and mixings are extracted.

To avoid numerical instability due to large cancellations between the
CP--even (CPE) and CP--odd (CPO) $\Sigma_D$ contributions to $m_{\nu}$
in Eq.~(\ref{neutralino34}), the combined contribution is obtained
using an analytic expansion in the 2 $\times$ 3 $\rpv$ matrices that
mix the Higgs with the sneutrinos in the 5 $\times$ 5 CPE and CPO
neutral scalar mass matrices. 
Both $\rpc$ and $\rpv$ corrections to the physical neutrino masses are
computed simultaneously, treating the charged sleptons and down--type Higgs as well
as the neutral scalars and Higgs particles on an equal footing, as in the
tadpole calculation. 

When calculating realistic inverted or quasi--degenerate neutrino mass
spectra, the mass eigenvalues are much larger than their differences,
so some fine tuning is expected.  For example, in
Ref.~\cite{Dreiner:2011ft}, the level of fine--tuning is estimated to
be of $\mathcal{O}(10^{-3})$ and $\mathcal{O}(10^{-4})$ for the
inverted and quasi--degenerate cases respectively.  In order to obtain
numerically stable results for neutrino masses and mixings for degenerate or
inverted hierarchies, we recommend setting the \code{TOLERANCE}
parameter to $10^{-5}$ and $10^{-6}$, respectively.

\section*{Acknowledgments}
This work has been partially supported by STFC, an IPPP associateship
and the Deutsche Telekom Stiftung. MH and CHK would like to thank
J.S. Kim for useful discussions.  CHK thanks the hospitality of Oxford
Rudolf Peierls Centre for Theoretical Physics for hospitality while
part of the work was carried out.

\appendix

\section{Running \SOFTSUSY}
\label{sec:run}

\SOFTSUSY~produces an executable called \code{softpoint.x}.  
Instructions on how to run
\code{softpoint.x} in the presence of $\rpv$ couplings are detailed in
Appendices A and B of Ref.~\cite{Allanach:2009bv}.  
Starting from version
\SOFTSUSY\code{-3.2}, output for both input options is compliant to the SLHA2
conventions, which are fully explained in Ref.~\cite{Allanach:2008qq}.
A sample SLHA2 file, \code{rpvHouchesInput}, may be used as input, and
\SOFTSUSY~run,  by the command 
\begin{verbatim}
./softpoint.x leshouches < rpvHouchesInput
\end{verbatim}
When any non--zero $\rpv$ couplings are set in either the command line 
or the $\rpv$ Les Houches input file, the program
automatically calls the neutrino mode instead of the original $\rpv$
mode. There is a new optional SLHA2 input parameter in the block SOFTSUSY
associated 
with the neutrino mode, which can be set in the following fashion: 
\begin{verbatim}
Block SOFTSUSY
    9   0.000000000e+00      # output uses normal hierarchy (=0.0) or inverted (=1.0)
\end{verbatim}
We also remind the reader that the quark mixing assumption can have a physical
effect given an assumed set of $\rpv$ couplings, and that the CKM mixing may be
set solely in the up or down quark sector via
\begin{verbatim}
Block SOFTSUSY
    2   1.000000000e+00      # quark mixing: none (=0), up (=1) or down (=2)
\end{verbatim}

\section{Sample Output \label{sec:output}}

The following is an example for one non--zero $\rpv$ coupling
$\lambda'_{322}(M_{GUT})=0.01$ at $M_{GUT}$ in an $\rpv$ mSUGRA
SPS1a--like~\cite{Allanach:2002nj} scenario:
\small
\begin{verbatim}
./softpoint.x sugra 100 250 -100 10 unified 1 lambdaP 3 2 2 0.01
\end{verbatim}
\normalsize The SLHA2 compliant output contains $\rpc$--conserving and
$\rpv$ MSSM information.  Results related to the neutrino
sector are displayed in the following blocks:
\begin{verbatim}
Block MASS   # Mass spectrum
#PDG code      mass              particle
        12     4.128113522451627e-50   # Mnu1(pole) normal hierarchy output
        14    -2.818848508640361e-42   # Mnu2(pole) normal hierarchy output
        16    -5.258228488951831e-06   # Mnu3(pole) normal hierarchy output
...
Block RVSNVEV Q= 4.670086691357357e+02 # sneutrino VEVs D 
     1   -2.115552831665844e-23   # SneutrinoVev_{1}
     2   -3.117652325269568e-18   # SneutrinoVev_{2}
     3   -7.695902798823816e-02   # SneutrinoVev_{3}
...
Block UPMNS Q= 4.670086691357357e+02 # neutrino mixing matrix:
  1  1     9.999999999518719e-01   # UPMNS_{11} matrix element
  1  2     9.811018586559509e-06   # UPMNS_{12} matrix element
  1  3     2.652785268618422e-20   # UPMNS_{13} matrix element
  2  1    -9.811018586559507e-06   # UPMNS_{21} matrix element
  2  2     9.999999999518718e-01   # UPMNS_{22} matrix element
  2  3     6.939913220612296e-17   # UPMNS_{23} matrix element
  3  1    -2.454643369143216e-20   # UPMNS_{31} matrix element
  3  2    -2.018886379347570e-16   # UPMNS_{32} matrix element
  3  3     1.000000000000000e+00   # UPMNS_{33} matrix element
\end{verbatim}
The block \code{MASS} includes the three pole neutrino masses
calculated in units of GeV, given the $\rpv$ model parameters.
Note that the corresponding particle data group (PDG) codes~\cite{pdg} are for
the 
neutrino flavour eigenstates. However, since we  
are dealing with massive neutrinos, we slightly adapt this definition such that the mass ordering of the
neutrinos is specified to be either normal or inverted hierarchy according to the value of the 
\code{invertedOutput} parameter.
In block \code{RVSNVEV}, the 1--loop sneutrino VEVs are displayed at the energy scale $Q$ at which 
the EWSB conditions are imposed ($M_S$).
 Finally, the block \code{UPMNS} displays the matrix elements of the matrix
$U_{\rm PMNS}$ in Eq.~\ref{eq:upmns}.
The neutrino spectrum predicted by this model point is in
contradiction with data, but in the next section, we calculate with
a parameter point that leads to neutrino mixings and masses that are agreement
with 2011 oscillation data
in a sample program. 

\section{Sample Program \label{sec:prog}}

In this section we present a sample main program, that illustrates a
calculation of neutrino masses and mixings given a set of $\rpv$
parameters at $M_{GUT}$.  This main program is included as the
\code{rpvneutmain.cpp}~file with the standard \SOFTSUSY~distribution
and performs the calculation assuming mSUGRA parameters $m_0=100$ GeV,
$M_{1/2}=500$ GeV, $A_0=932.9$ GeV, $\tan \beta=20$ and $\mu>0$, with
non--zero $\rpv$ parameters $\lambda'_{111}(M_{GUT})=0.039$,
$\lambda'_{211}(M_{GUT})=-0.016$, $\lambda'_{311}(M_{GUT})=0.018$,
$\lambda'_{133}(M_{GUT})=3.0\times10^{-5}$,
$\lambda'_{233}(M_{GUT})=3.0\times 10^{-5}$,
$\lambda'_{333}(M_{GUT})=-3.5\times10^{-5}$.  

The sample program has the following form: 
\small
\begin{verbatim}
#include <rpvmain.h>

int main() {
  /// Sets up exception handling
  signal(SIGFPE, FPE_ExceptionHandler); 

  /// MIXING=0: diagonal Yukawa matrices, MIXING=1: CKM-mixing in up-sector, 
  /// MIXING=2: CKM-mixing in down-sector
  MIXING = 1; 

  /// Apply SUSY breaking conditions at GUT scale, where g_1=g_2
  bool gaugeUnification = true;

  /// Sets format of output: 3 decimal places
  outputCharacteristics(3);
  
  /// "try" catches errors in main program and prints them out
  try {
    QedQcd oneset;      ///< See "lowe.h" for default parameter definitions 
    oneset.toMz();      ///< Runs SM fermion masses to MZ

    /// Guess at GUT scale
    double mxGuess = 2.e16;
    
    /// Close to scenario IH S2 from arXiv 1106.4338
    int sgnMu = 1;
    double tanb = 20., a0 = 912.3, m12 = 500., m0 = 100.;

    /// Define RpvNeutrino object    
    RpvNeutrino kw;

    /// Set the GUT scale RPV SUSY couplings
    kw.setLamPrime(1, 1, 1,  0.040);
    kw.setLamPrime(2, 1, 1, -0.018);
    kw.setLamPrime(3, 1, 1,  0.019);	
    kw.setLamPrime(1, 3, 3,  3.3e-5);
    kw.setLamPrime(2, 3, 3,  3.2e-5);
    kw.setLamPrime(3, 3, 3, -3.5e-5);

    /// Store inputs into one vector
    DoubleVector pars(3); pars(1) = m0; pars(2) = m12; pars(3) = a0;

    /// Outputs the RPV couplings required into the vector pars used by lowOrg
    kw.rpvDisplay(pars);
    /// Makes sure the neutrino mass ordering will be as expected in inverted
    /// hierarchy output. If required, must be set before lowOrg is called 
    kw.setInvertedOutput();
    /// For inverted output, because of large cancellations, one requires the
    /// SUSY spectrum calculation to be performed to a high fractional
    /// accuracy
    TOLERANCE = 1.0e-6;

    /// Main driver routine: do the calculation
    double mgut = kw.lowOrg(rpvSugraBcs, mxGuess, pars, sgnMu, tanb, oneset, gaugeUnification);

    /// Output the results in SLHA2 format
    double qMax = 0.;  char * modelIdent = (char *)"sugra"; 
    int numPoints = 1; bool altEwsb = false;
    kw.lesHouchesAccordOutput(cout, modelIdent, pars, sgnMu, tanb, qMax, numPoints, mgut, altEwsb);
  }
  catch(const string & a) {
    cout << a; exit(-1);
  }
  catch(const char *a) {
    printf("%s", a); exit(-1);
  }
}
\end{verbatim}
\normalsize The structure of the main file above is as follows.  After
including a header file, an exception handler is set up, followed by
specifying a (sub--) set of global variables, all of which are
described in the $\rpc$ manual~\cite{Allanach:2001kg}.  These include
the \code{MIXING}~switch, which determines how any quark mixing is
implemented, the
\code{gaugeUnification} switch, which determines \code{mGutGuess}~as
the scale of $M_{GUT}$ of electroweak gauge unification. The
switch 
 \code{outputCharacteristics}, which specifies the output
accuracy, is then set.

The running masses of the SM fermions and the QED and QCD gauge
couplings are determined at $M_Z$ from data with the method
\code{toMz}.  In order for \SOFTSUSY~to determine $M_{GUT}$, it
requires an initial guess, which must be supplied as the initial value
of the variable \code{mxGuess} (in GeV), and is later over--written by
the program with a more accurate calculated value\footnote{If the user wishes
  to provide this, $2 \times 
  10^{16}$ GeV is a good initial guess for $M_{GUT}$.}.

Next, the mSUGRA parameters ${\rm
  sgn}\mu=$\code{sgnMu}, $\tan \beta=$\code{tanb},
$A_0/$GeV$=$\code{a0}, $M_{1/2}/$GeV$=$\code{m12}  and
$m_0/$GeV$=$\code{m0} are defined.  This is then followed by the instantiation
of the \code{RpvNeutrino} object \code{kw} and the assignment of the
$\rpv$ parameters \code{kw.setLamPrime(i,j,k,val)} for
$\lambda'_{ijk}=$\code{val}.  In the iterative \SOFTSUSY~algorithm the
parameters in the \code{RpvNeutrino} object change due to the RGE
running.  The \code{pars} vector is needed to keep track of the
boundary conditions set at $M_{GUT}$.  These boundary conditions are
re--set in every iteration at $M_{GUT}$ from the unchanged
\code{DoubleVector pars} parameters.

We do not fill the other 102 $\rpv$ entries of \code{pars}
explicitly. This would be tedious and an additional source of
potential bugs. Instead, we fill the \code{RpvSoftsusy} object itself
using the \code{setLamPrime} method in this example.  We use the
\code{rpvDisplay} method: this fills the \code{pars} vector
automatically with what was set already inside the \code{RpvNeutrino}
object, while leaving the first nine entries in the vector unchanged.
The \code{rpvDisplay} method automatically changes the length of
\code{pars} appropriately.  

The neutrino masses may be presented in normal (default) and inverted
mass orderings.  This is controlled by the method
\code{setInvertedOutput} which must be invoked before the actual
\SOFTSUSY~main driving method \code{lowOrg} is called.  When
\code{lowOrg} is called, the first argument specifies the type of
boundary condition (currently \code{rpvSugraBcs}), which assumes that
\code{pars}~has already been prepared by using the \code{rpvDisplay}
object.  This is followed by a print to standard output of the model
parameters at $M_Z$ and the physical parameters SLHA2 compliant format.
Finally, the 
\code{catch} commands print any errors produced by the code.

The output is in standard SLHA2 format, including
neutrino masses in GeV units, neutrino mass 
ordering (inverted hierarchy 
in this example) and the PMNS mixing matrix: 
\begin{verbatim}
[ ... ]
Block MASS                      # Mass spectrum
# PDG code     mass             particle
[ ... ]
        12     4.65793809e-11   # Mnu1 inverted hierarchy output
        14     4.70348794e-11   # Mnu2 inverted hierarchy output
        16     1.20158889e-14   # Mnu3 inverted hierarchy output
[ ... ]
Block UPMNS Q= 9.11876000e+01 # neutrino mixing matrix (inverted  hierarchy)
  1  1     8.61178721e-01   # UPMNS_{11} matrix element
  1  2     5.08244716e-01   # UPMNS_{12} matrix element
  1  3     7.64979577e-03   # UPMNS_{13} matrix element
  2  1    -3.29127098e-01   # UPMNS_{21} matrix element
  2  2     5.69021210e-01   # UPMNS_{22} matrix element
  2  3    -7.53584909e-01   # UPMNS_{23} matrix element
  3  1    -3.87358444e-01   # UPMNS_{31} matrix element
  3  2     6.46453533e-01   # UPMNS_{32} matrix element
  3  3     6.57306066e-01   # UPMNS_{33} matrix element
[ ... ]
\end{verbatim}

\section{The \code{RpvNeutrino}~class\label{sec:objects}}

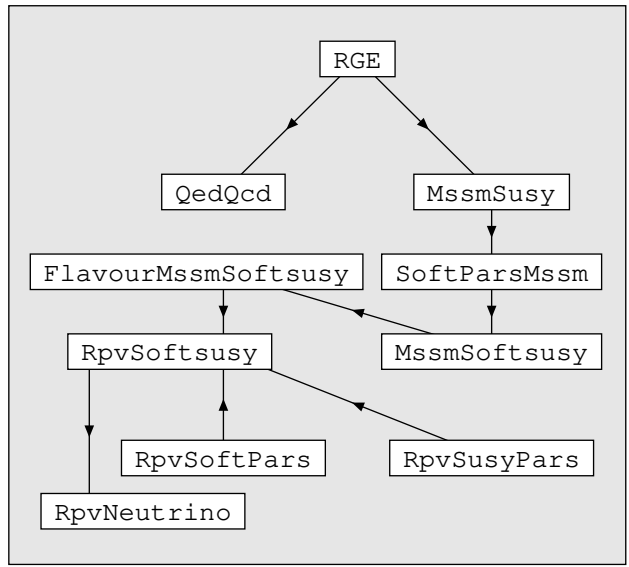
\begin{figure}
\begin{center}
\begin{picture}(215,200)(-30,0)
\GBox(200,200)(-30,-10){0.9}
\ArrowLine(100,180)(50,130)
\ArrowLine(100,180)(150,130)
\ArrowLine(150,130)(150,100)
\ArrowLine(150,100)(150,70)
\ArrowLine(150,70)(50,100)
\ArrowLine(50,100)(50,70)
\ArrowLine(50,30)(50,70)
\ArrowLine(150,30)(50,70)
\ArrowLine(0,70)(0,10)
\SetPFont{Teletype}{10}
\BText(100,180){RGE}
\BText(50,130){QedQcd}
\BText(150,130){MssmSusy}
\BText(50,30){RpvSoftPars}
\BText(40,100){FlavourMssmSoftsusy}
\BText(30,70){RpvSoftsusy}
\BText(150,100){SoftParsMssm}
\BText(150,70){MssmSoftsusy}
\BText(150,30){RpvSusyPars}
\BText(20,10){RpvNeutrino}
\end{picture}
\end{center}
\caption{Heuristic high-level object structure of \SOFTSUSY-3.2. Inheritance is
displayed by the direction of the arrows. \label{fig:objstruc}}
\end{figure}

We now go on to sketch the class \code{RpvNeutrino}.  This class
publicly inherits from the class \code{RpvSoftsusy}, \textit{cf.} Fig.~\ref{fig:objstruc}. 
The \code{RpvSoftsusy} methods \code{physical},  \code{calculateSneutrinoVevs} 
and  \code{doCalcTadpole1oneLoop},  \code{doCalcTadpole2oneLoop} are overloaded  in \code{RpvNeutrino} in order to 
include 1--loop neutrino masses, sneutrino VEVs and tadpoles.
The data and methods in the
\code{RpvNeutrino} class deemed of possible importance for prospective
users are presented in Table~\ref{tab:RpvNeutrino}.
Note that different quark mixing 
assumptions can significantly 
affect the predictions of the neutrino masses. The \code{MIXING} parameter is
implemented   in the neutrino mode in the same fashion as in
Ref.~\cite{Allanach:2001kg}, and can take values \code{0, 1, 2}~for no mixing,
up quark mixing and down quark mixing, respectively.

\begin{table}\begin{center}
\begin{tabular}{lll}
  data variable & & methods \\ \hline
  \code{\small DoubleVector physNuMasses} & physical neutrino masses&
  \code{\small displayPhysNuMasses} \\ 
  $m_{\nu_i={1,2,3}}/$GeV  & & \code{\small setPhysNuMasses}\\ \hline
  \code{\small DoubleMatrix uPmns} & PMNS mixing matrix & \code{\small
    displayUpmns} \\
  $U_{\rm PMNS}$ (3 $\times$ 3) &  & \code{\small setUpmns}\\ \hline
    \code{\small DoubleMatrix physNeutMix} & physical neutral fermion mixing&
  \code{\small displayPhysNeutMix} \\ 
  $O$ (7 $\times$ 7) & & \code{\small setPhysNeutMix}\\ \hline
  \code{\small bool invertedOutput} &neutrino mass ordering&
  \code{\small displayInvertedOutput} \\
  & & \code{\small setInvertedOutput} \\
  & & \code{\small setNormalOutput} \\ \hline
\code{\small double theta12, theta23} & 
PDG parameterisation  & \code{\small displayThetaCkm12, displayThetaCkm23} \\
\code{\small double theta13, deltaCkm} & 
of CKM angles  & \code{\small displayThetaCkm13, displayDeltaCkm} \\
$\theta_{12}, \theta_{23}, \theta_{13}, \delta$/radians & 
  & \code{\small setThetaCkm12, setThetaCkm23} \\
 & & \code{\small setThetaCkm13, setDeltaCkm} \\ \hline
\code{\small DoubleVector CPEmasses} & tree-level CP even neutral & \code{\small displayCPEMasses} \\
$(m_{\varphi^0_+})_{i=1\ldots 5}$ & scalar masses & \code{\small setCPEMasses} \\ \hline
\code{\small DoubleVector CPOmasses} & tree-level CP odd neutral & \code{\small displayCPOMasses} \\
$(m_{\varphi^0_-})_{i=1\ldots 5}$ & scalar masses & \code{\small setCPOMasses} \\ \hline
\code{\small DoubleMatrix CPEscalars} & tree-level CP even neutral &
\code{\small displayCPEscalars} \\
${\mathcal M}^2_{\varphi^0_+}$ $(5 \times 5) $ & scalar mass$^2$ matrix & \code{\small setCPEscalars} \\ \hline
\code{\small DoubleMatrix CPOscalars} & tree-level CP odd neutral &
\code{\small displayCPOscalars} \\
${\mathcal M}^2_{\varphi^0_-}(5 \times 5) $ & scalar mass$^2$ matrix & \code{\small setCPOscalars} \\ \hline
\code{\small DoubleMatrix CPEscalarMixing} & tree-level CP even neutral &
\code{\small displayCPEscalarMixing} \\
$O_+$ $(5 \times 5) $ & scalar mixing matrix & \code{\small setCPEscalarMixing} \\ \hline
\code{\small DoubleMatrix CPOscalarMixing} & tree-level CP odd neutral &
\code{\small displayCPOscalarMixing} \\
$O_-$ $(5 \times 5) $ & scalar mixing matrix & \code{\small setCPOscalarMixing} \\ \hline

\end{tabular}\end{center}
\caption{\code{\small RpvNeutrino} class. We display the important data
contained within the object, along with ways of accessing and setting their
values.  For the definitions of the methods, see the file \code{rpvneut.h}~in
the \SOFTSUSY~distribution. 
\label{tab:RpvNeutrino}}
\end{table}

\end{document}